\documentclass[a4paper]{ESLAB2005}
\usepackage{epsfig}
\begin{document}

\title{The magnetism of the solar interior for a complete MHD solar vision}

\author{S. Turck-Chi\`eze\inst{1}, T. Appourchaux\inst{2}, 
J. Ballot\inst{1}, G. Berthomieu\inst{3}, P. Boumier\inst{2}, A. S. Brun\inst{1},   
A. Cacciani\inst{4},
J. Christensen-Dalsgaard\inst{5}, T. Corbard\inst{3}, S. Couvidat\inst{6}, 
A. Eff-Darwich\inst{7}, B. Dintrans\inst{8}, E. Fossat\inst{9}, R. A. Garcia\inst{1}, 
B. Gelly\inst{10}, L. Gizon\inst{11}, D. Gough\inst{12}, A. Jimenez\inst{7}, 
S. Jimenez-Reyes\inst{7}, A. Kosovishev\inst{6}, P. Lambert\inst{1}, 
I. Lopes\inst{13},  M. Martic\inst{14}, S. Mathis\inst{15}, N. Meunier\inst{8}, P. A. Nghiem\inst{1},
P. Palle\inst{7}, L. Piau\inst{16}, J. Provost\inst{3}, M. Rieutord\inst{8}, 
J. M. Robillot\inst{17}, T. Roudier\inst{8}, I. Roxburgh\inst{18}, J. P. Rozelot\inst{13},
S. Solanki\inst{11}, S. Talon\inst{19}, M. Thompson\inst{20}, G. Thuillier\inst{14}, 
S. Vauclair\inst{8}, J. P. Zahn\inst{15}}
\institute{SAp/DAPNIA/CEA CE Saclay, 91191 Gif sur Yvette, France
\and Institut d'Astrophysique Spatiale, 91405 Orsay Cedex, France
\and UMR CNRS 6202, Observatoire de la C\^ote d'Azur, BP 4229, 06304 Nice Cedex 4, France
\and Dipartimento di Fisica, Universit\`a degli Studi di Roma `La Sapienza', Piazzale Aldo Moro 2, I-00185 Roma, Italy
\and Institut for Fysik og Astronomi, Aarhus Universitet, DK-8000 Aarhus C, Denmark
\and W. W. Hansen Experimental Physics Laboratory, Stanford University, Stanford, CA 94305, United States
\and Instituto de Astrofísica de Canarias, E-38205 La Laguna, Spain
\and Observatoire Midi Pyr\'en\'ee et Laboratoire d'Astrophysique de Tarbes, France
\and D\'epartement d'Astrophysique, UMR 6525, Universit\'e de Nice-Sophia Antipolis, 06108 Nice Cedex 2, France
\and THEMIS S.L., 38205, La Laguna, Tenerife, Spain
\and  Max Planck Lindau, Germany
\and Institute of Astronomy, University of Cambridge, Cambridge CB3 0HA, England
\and Istituto Superior T\'ecnico, Lisboa, Portugal
\and Service d'A\'eronomie du CNRS, 91370 Verri\`eres-le-Buisson, France
\and Observatoire de Paris, LUTH Meudon, France
\and Dept of Astron. \& Astroph., Chicago, United States 
\and Observatoire de l'Universit\'e Bordeaux 1, BP 89, 33270 Floirac, France
\and Astronomy Unit, Queen Mary, University of London, Mile End Road, London E1 4NS, United Kingdom 
\and D\'epartement de Physique, Universit\'e de Montr\'eal, Montr\'eal PQ H3C 3J7, Canada  Canada
\and School of Science \& Mathematics, Sheffield Hallam University, Sheffield S1 1WB, United Kingdom}
\maketitle 
\vspace{-0.4mm}
\begin{abstract}
\vspace{-0.3mm}
The solar magnetism is no more considered as a purely superficial phenomenon.
The SoHO community has shown that the length of the solar cycle depends on the 
transition region between radiation and convection. 
Nevertheless, the internal solar (stellar) magnetism stays poorly known. 
Starting in 2008, the American instrument HMI/SDO 
and the European 
microsatellite PICARD will enrich our view of the Sun-Earth relationship.
Thus obtaining a complete MHD solar picture is a clear objective for the next decades and it requires 
complementary observations 
of the dynamics of the radiative zone. 
For that ambitious goal, space prototypes are being developed to improve gravity mode detection.
The Sun is unique to progress on the topology of deep internal magnetic fields
and to understand the
complex mechanisms which provoke  photospheric and coronal magnetic changes and possible longer cycles 
important for human life. 
We  propose the following roadmap in Europe to contribute to this "impressive" revolution in Astronomy 
and in our Sun-Earth relationship:
SoHO (1995-2007), PICARD (2008-2010), 
DynaMICS (2009-2017) in parallel to SDO (2008-2017) then a world-class mission 
located at the L1 orbit or above the solar poles.

\end{abstract}

\vspace{-1.2cm}
\section{Introduction}
Magnetic field plays a fundamental role in the Universe. Most of the celestrial bodies are rotating and dynamo 
action produces or amplifies magnetic field
in stars, planets and galaxies. But different kinds of dynamo must be studied,   
it is now believed that the initial conditions are important to understand 
stellar and galactic magnetic fields  
(Rudiger \& Hollerbach 2004). We need  to determine magnetic field saturation growth 
and which magnetic configurations are stable (Spruit 2002). 
We have today a very poor information on this difficult ingredient 
of Astrophysics and  it is partly why it is absent 
from most of the present simulations.  Nevertheless, its role is extremely crucial 
in star formation, to reproduce the extension of the convective zone in young stars 
(D'Antona et al. 1998, Piau \& Turck-Chi\`eze 2002), 
and probably to model massive presupernovae... 
Moreover, it begins to be an important ingredient to 
precisely estimate the Sun-Earth 
relationship and to predict coming and possibly different solar cycle(s).

\begin{figure*}[t!]
\centering
\includegraphics[width=.34\textwidth, height=0.34\textheight,angle=90]{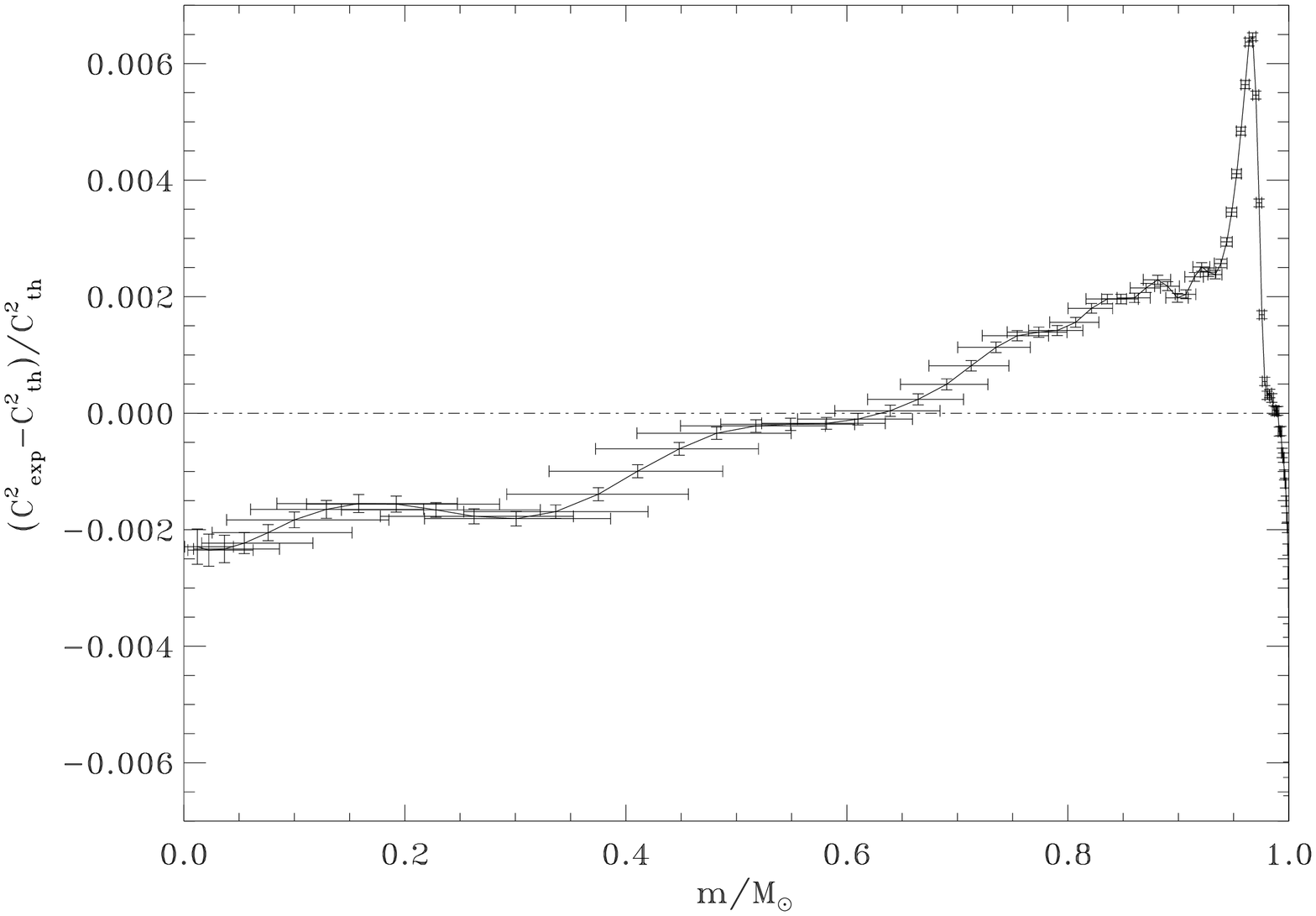}
\includegraphics[width=.42\textwidth,height=0.26\textheight]{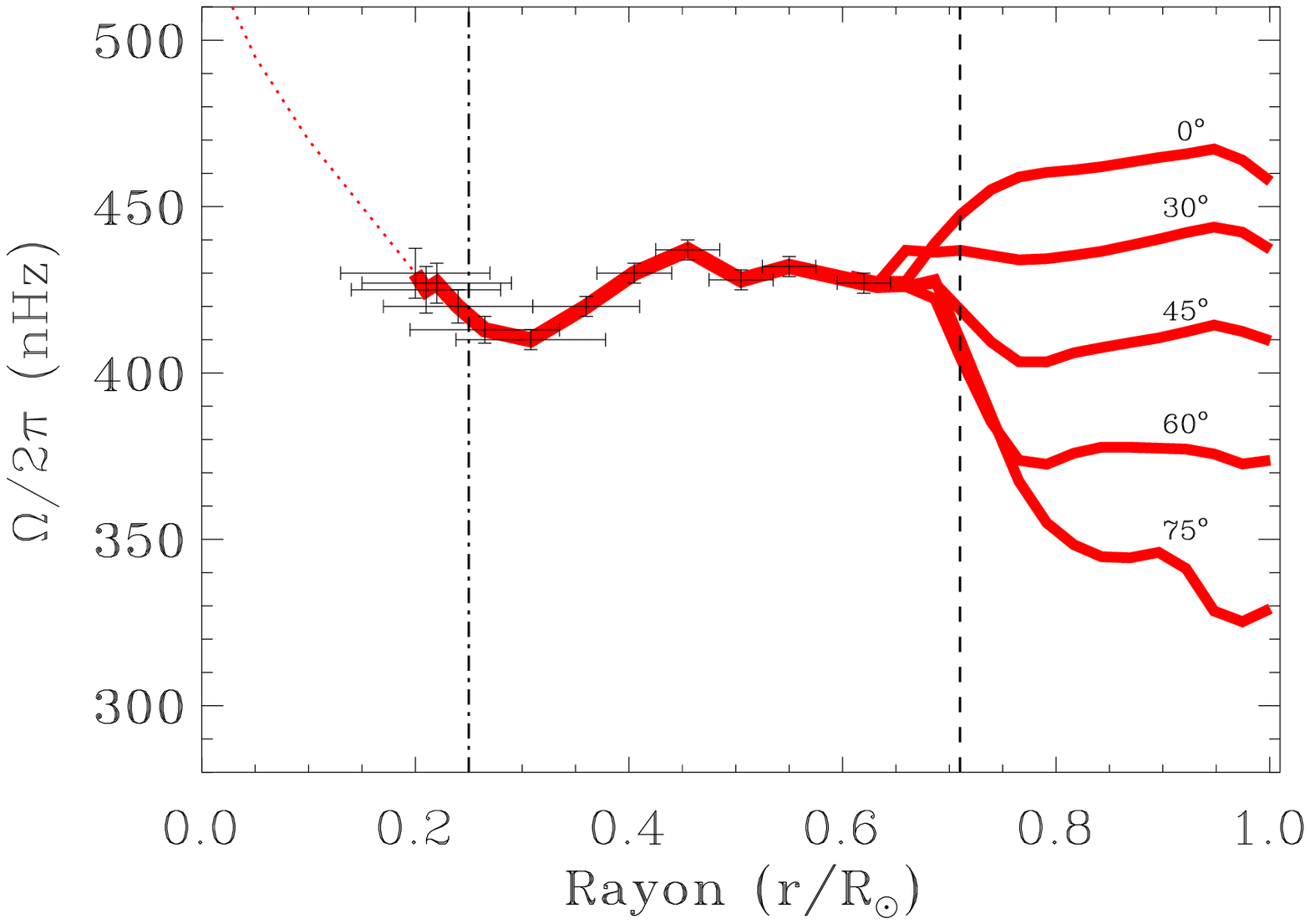}
\caption{\small { Left a) Relative squared sound speed difference between a standard model 
(Turck-Chi\`eze et al., 2001a) and the Sun, using GOLF+MDI observations (Bertello et al, 2000), versus 
the fraction mass.
One notices a spatial resolution of about $\pm 6\%$ in mass and an absolute determination of the sound speed 
of the order of 10$^{-4}$ (Turck-Chi\`eze et al, 2001b). The transition between radiation and convection in located at about 98\%
in mass fraction. Right b) Solar rotation profile deduced from acoustic mode observations done by GOLF/MDI aboard SoHO 
(Kosovichev et al 1997, Couvidat et al. 2003a, Garcia et al. 2004).}}
         \label{result}
\end{figure*}

Nowadays 2D-3D simulations are quickly developing, although the capabilities of the 
more advanced computers are still limited for realistic simulations. Observational constraints 
are extremely important to orientate this effort; the Sun is the only star for which
detailed internal dynamical phenomena can be observed. Thus new solar missions are crucial for next decades, 
in parallel to the development of asteroseismology and exoplanet finding.

The classical stellar structure equations include a rich microphysics and indirect rotation effects
but they still ignore magnetic field and magnetic field transport effects. 
3D Magneto Hydro Dynamics simulations tentatively describe the internal motions for portions of stars with a limited 
description of  microphysics and limited timescale. We need observational internal constraints on the dynamical 
processes from the core 
to the surface and beyond to build a realistic complete MHD description of stars 
and their environmental planets. In this respect, we cannot ignore the dynamics of the solar 
radiative zone which represents 98\% of 
the solar mass (Fig 1a). Direct and indirect informations on internal magnetic fields will come from new space solar 
observations.

The SoHO mission has been successful to induce the first step towards this complete MHD vision of our Sun; 
the most important results are rapidly summarized in section 2. As the already programmed missions 
will largely improve 
the convective zone, photospheric phenomena and corona, we will mainly focus on the radiative region
 which needs continuous activities during the next two decades.
The dynamical processes connected to rotation, 
magnetic field and internal waves are introduced in section 3.
The great interest for complete MHD simulations is demonstrated in section 4 and the 
open questions for observers are recalled in section 5. 
Gravity modes are the extremely promising probes for 
the radiative zone (section 6): they are
actively searched with the SoHO satellite and stimulate coordinated efforts 
to improve the signal/noise ratio in the next generation of instruments. 
The last section 7 is devoted to the missions which must fly during the next two decades.
Two techniques are presently studied and must be launched quickly to be validated:
a prototype of a space Resonant Spectrometer will be 
installed in Tenerife in 2006 and another prototype of a Magnetic Optical Filter at Dome C.
In confronting the open questions, the present scheduled missions and the improved techniques, we build 
a roadmap for Europe which converges to our objective: a complete MHD solar vision at the horizon 2018
 with a world-class 
mission at the Lagrangian point or above the solar poles.  

\section{Fifteen years of seismic solar investigation}
Ground networks and the three space seismic instruments aboard SoHO have dramatically changed our stellar vision.
From a  static picture, we have slightly evolved 
towards a more dynamical view, thanks to the detection of most of the acoustic modes. 
Some theoretical concepts 
as meridional circulations or latitudinal flows have become reality in the upper convective zone. 
The slow migration from high latitudes to the equator of the superficial spots
have been followed in time and in depth during the last half cycle showing migrating bands 
rotating faster or slower. The torsional oscillations corresponding 
to time variations of the solar rotation (Howe et al. 2000, Antia \& Basu 2000, Vorontsov et al. 2002) 
are nowadays confronted to non linear axisymmetric mean-field dynamo model (Vorontsov et al. 2003).
The solar rotation profile (Fig 1b) has been nowadays properly established down to the limit of the nuclear core
(Kosovichev et al, 1997, Chaplin et al. 2001, Couvidat et al. 2003a, Garcia et al. 2003). In contradicting 
all the theoretical predictions
it represents today a strong challenging information for theoreticians (see sections 3 and 4). The thin width 
of the tachocline 
(transition region from differential rotation to flat rotation) of about 5\% solar radius, plays 
a strong role in the understanding of the solar dynamo (Brummell, Cline \& Cattaneo 2002 and section 4). 
The sound speed profile 
is now very precisely established (Fig 1a, Turck-Chi\`eze et al. 2001a,b; Couvidat et al. 2003b)) down to 6\% R$_\odot$ but this information is extracted from the low frequency 
acoustic modes, which are less influenced by the turbulence and the magnetic field of the superficial layers. However the spatial resolution is not 
sufficiently good to get a proper determination of the density profile. 

Despite these unprecedent advances useful not only for the Sun but also for all the stars,
 magnetic field is only directly constrained just below the photosphere, 
elsewhere only upper limits are presently available.

\section{The internal radiative solar (stellar) dynamics}
Considering that the American project SDO will largely improve our knowledge of the convective zone, 
we will mainly concentrate in this presentation on the radiative zone. The main stellar results 
come from classical stellar evolution codes which suppose that the star is spherical and is not 
influenced by magnetic field.

In these models, radiation zones are treated as stable regions, 
where no motion occurs, apart from rotation in some specific cases. 
But various observations (i.e. surface abundances, 
sound speed profile) show that these zones are the seat of mild mixing. The most likely cause of 
such mixing is the (differential) rotation, which drives a large scale meridional circulation and 
gives probably rise to shear turbulence. This process has been described in a self-consistent way, 
namely taking into account the transport of angular momentum which modifies the rotation profile 
(Zahn 1992, Maeder \& Zahn 1998). Series of models have been built which include these processes, 
called rotational mixing (Talon et al. 1997,  Meynet \& Maeder 2000, 2005), and for massive stars, 
the agreement with observations is largely improved. However, rotational mixing alone cannot explain the 
almost uniform rotation of the radiative interior of the Sun, as revealed through helioseismology 
(Couvidat et al. 2003a, Thompson et al. 2003). Therefore, another more powerful process is probably operating, 
at least in slow rotators. 

\begin{figure}[h!]
\centering
\resizebox{8cm}{!}{\includegraphics{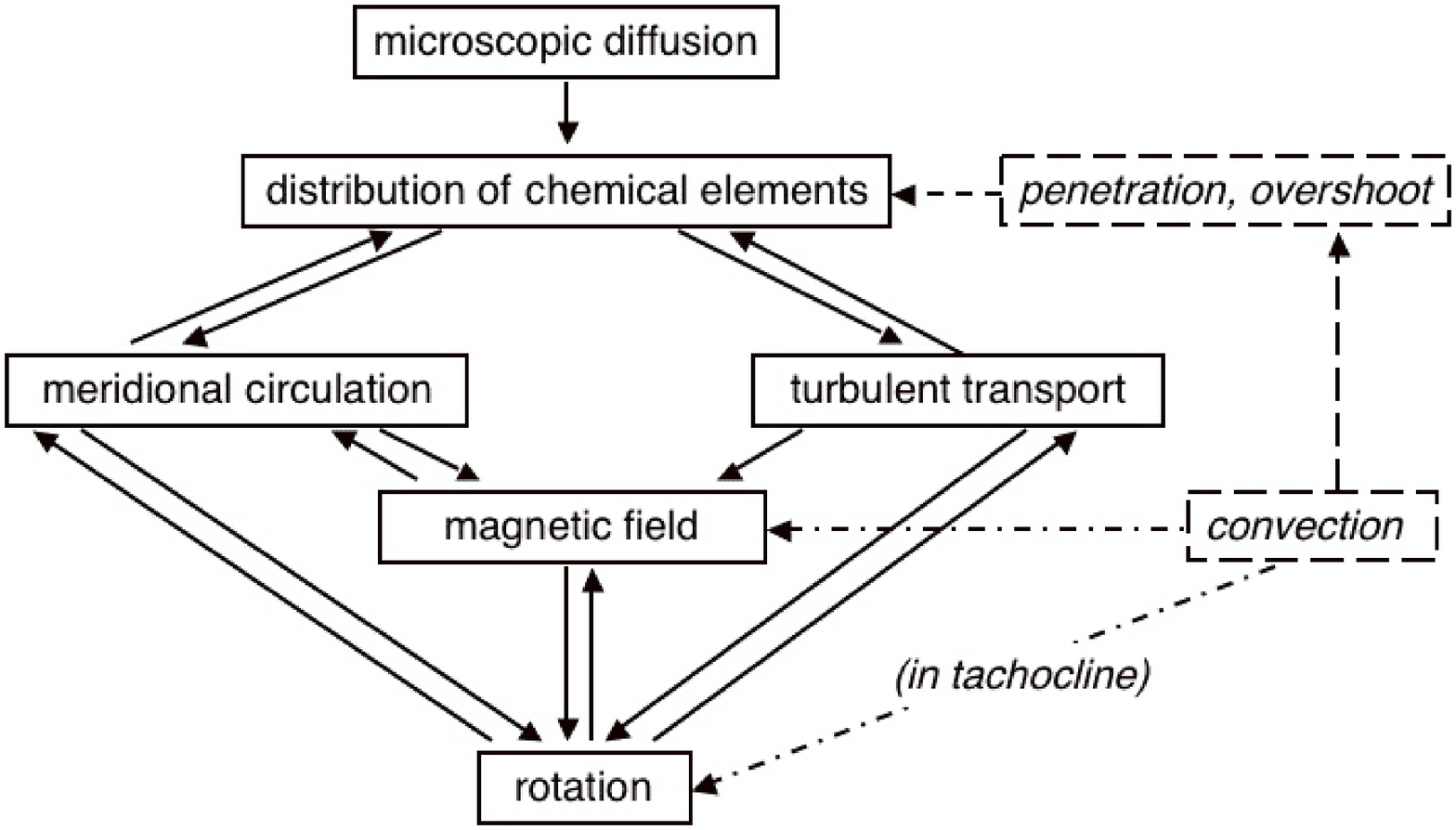}}
\caption{Transport and mixing processes in stellar interiors with magnetic field: a highly non-linear problem.}
\label{diagram}
\end{figure}

Two ideas have been proposed to explain the quasi-flat solar rotation profile, namely magnetic 
torquing (e.g. Charbonneau \& MacGregor 1993, Gough \& McIntyre 1998, Garaud 2002) and momentum redistribution by internal 
gravity waves (Talon et al.
2002). Much theoretical work is currently done in order to take into account 
these physical mechanisms. A new theoretical frame suggests to treat simultaneously the 
core of the radiation zones and the tachoclines (Mathis \& Zahn 2004), in order to capture the mixing 
which occurs there (Spiegel \& Zahn 1992, Brun et al. 2002). Also better prescriptions now exist to 
describe the anisotropic turbulent transport (Maeder 2003, Mathis et al. 2004) as well as equations 
to consistently introduce the effect of a fossil magnetic field (Mathis \& Zahn 2005) as was attempted 
previously by Mestel et al.
(1988). The only way to treat the unstationary problem in a consistent way is to
 take into account the 
advection of the field by the meridional circulation, the production of toroidal field (by shearing the 
poloidal field through the differential rotation), but also the feedback of the magnetic field on the rotation 
profile and on the circulation, through the Lorentz force (see figure 2). Presently, work is in progress to implement these 
equations in stellar evolution codes. 

In the case of internal gravity waves, Talon \& Charbonnel (2005) recently developed a formalism 
to describe the evolution of the rotation profile over evolutionary time-scale, by modeling 
the effect of the rapidly oscillating shear layer generated by waves just at the base of the 
surface convection zone and originally described by Ringot (1998), see also Kumar, Talon \& Zahn (1999). 
For a given wave flux generated by turbulence in the convection zone, it is now possible to predict 
the state of radial differential rotation at the solar age, which, for conservative estimates of the 
wave flux, is rather flat (Charbonnel \& Talon 2005). Details regarding the core rotation rate are 
crucial to distinguish between the two mechanism exposed here.

\section{Radiative Interior MagnetoHydrodynamics}

Many models and simulations have been developed to understand the 
complex interactions between convection, turbulence, rotation and magnetic field
seen at the solar surface and in its upper atmosphere. Most of these studies have
been carried out in the context of the solar dynamo puzzle or in an attempt to model the
intricate solar atmosphere, but very few to study the dynamics of the solar radiative interior.
It is now time to develop a global model of the solar dynamics and 
magnetism from deep within its radiative interior up to its corona and extended wind.

\begin{figure*}[!ht]
\center
\includegraphics[width=180mm]{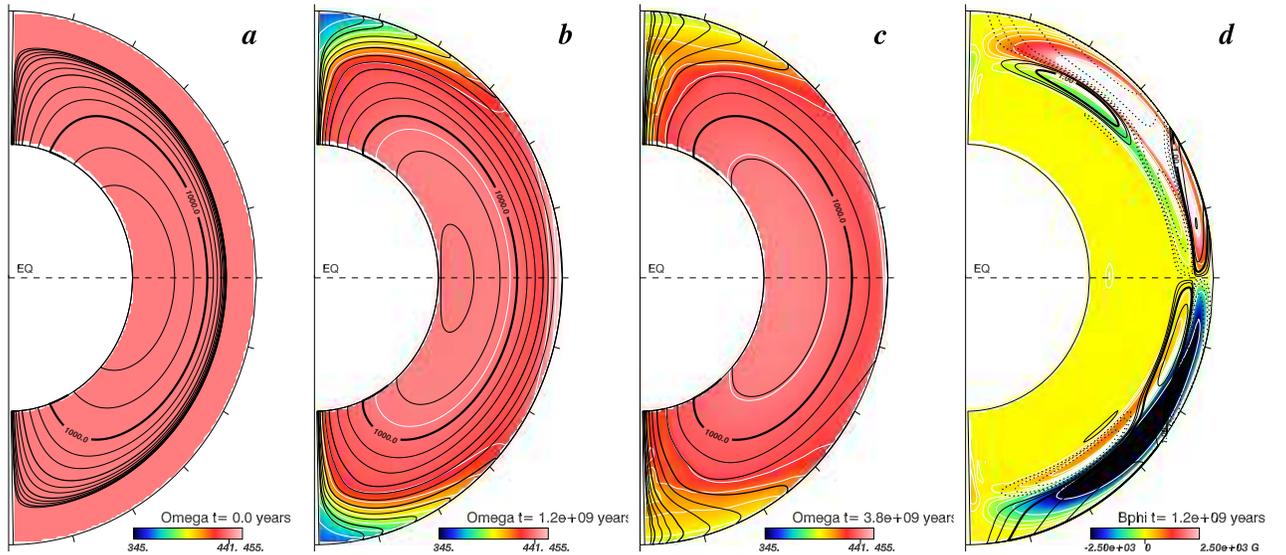}
\caption[]{\label{figKEMETac} a-c) Temporal evolution of a dipolar magnetic field in the
presence of rotation and shear. We note the connection of the field lines in the imposed
top shear and the resulting Ferraro's law of isorotation at high latitude.
d) The associated toroidal field and meridional circulation at t = 1.2 Gyr. The presence of
a large scale shear helps the magnetic field to reach a more stable mixed poloidal/toroidal 
configuration (from Brun \& Zahn 2005).}    
\end{figure*}
Indeed it is key to realize that 98\% of the solar mass is contained in the solar radiative zone
and that complex magnetohydrodynamical processes that continuously shape the upper atmosphere activity 
take place rather deep in the Sun. Understanding the dynamical properties of the deep
interior is thus one of the major challenges that solar physicists have to address/confront by developing
news observational and numerical tools.
Major questions remain to be solved such as:
\begin{itemize}
\item the dynamical influence of rotation and magnetic field
\item the origin of the internal solid body rotation 
\item the main agents responsible for the redistribution of the angular momentum
\item the topology, strength and influence of a fossil field
\item the role of the g-modes and progressive internal waves in the overall internal
dynamics
\item the nature of nonlinear interactions between the dynamo and fossil fields
\item the presence of large scale flows, their amplitude and shape and their mixing properties
\item the presence of hydrodynamical or magnetohydrodynamical instabilities and their coupling
\item the nonlinear interaction and coupling between the radiative and convection zone
\item the profiles, degree of ionization of the chemical species and the transport of photons in dense plasma
\item the role and influence of the tachocline on the internal structure and composition
\end{itemize}

The interaction of a stably stratified zone with either 
rotation or magnetic fields have been addressed for quite a long time: Von Zeipel 1924, Sweet 1950, Tayler 1973, 
Pitts \& Tayler 1985, 
Mestel \& Weiss 1987, Moss et al. 1990, Zahn 1992, Spruit 2002 and references therein.
They note that the dynamical properties of 
rotating magnetized radiative interiors are much more complicated than expected, and
give rise to a large range of both HD or MHD instabilities.

The discovery of the so-called tachocline (see section 2) has put forward the urgent need for the solar 
community to better understand the MHD of the radiative interior and of the tachocline and their
coupling to the upper layers. It seems now well established, that the tachocline plays a crucial role in the 
solar dynamo since it is most likely the layer where the mean toroidal magnetic field, 
thought to be at the origin of the surface sunspots and butterfly diagram, 
is streched, amplified (by at least a factor of 100) and stored until it
becomes magnetically buoyant (Parker 1993, Caligari et al. 1998, Fisher et al. 2000, Rempel 2003). 
However little is known about the dynamics of the solar tachocline: is it turbulent or laminar, what 
type of circulations are present, what is the dynamical influence of the magnetic fields, why is 
it so thin extending at most over 5\% of the solar radius? (Spiegel \& Zahn, 1992) were the 
first to address directly some of those questions but in the purely hydrodynamical context. 
They showed that if no processes were present to oppose its radiative spread, the solar tachocline 
would extend over 30\% of the solar radius after 4.6 Gyr, in complete contradiction with current helioseismic 
inversions (Corbard et al. 1999). They demonstrate that the anisotropic turbulence could hinder the 
spread of the solar tachocline to only few \% of the solar radius. 
Elliott (1997) confirmed their results numerically with a 2--D axisymmetric
hydrodynamic code. Miesch (2003) using a thin layer version of the ASH code 
(Clune et al. 1999, Brun et al. 2004), showed that 
the coupling between randomly forced turbulence and an imposed shear flow gives rise 
to Reynolds stresses that transport angular momentum such as to reduce the shear. 
However several authors (Rudiger \& Kitchatinov 1997; Gough \& Mc Intyre 1998; 
MacGregor \& Charbonneau 1999; Garaud 2002) have proposed that the magnetic torques 
exerted by a weak internal fossil magnetic field (if it exists) could oppose the inward 
thermal hyperdiffusion (or viscous diffusion when thermal effect are neglected) 
of the solar tachocline. Such models favor a slow, rather laminar version of
the tachocline. Gilman and collaborators (Dikpati, Cally \& Gilman 2004 and 
references therein) developed a series of models that showed that the tachocline could become unstable through
magnetic instability of toroidal structures embedded within it, resulting
in a latitudinal angular momentum transport that suppress the shear and limit its inward 
diffusion. Forgacs-Dajka (2004) considered the effect of an oscillating dynamo field and found that it
could also suppress the spread of the solar tachocline. This increasing number of solar tachocline
models demonstrates how important it is to characterize the dynamical properties 
of the solar radiative interior. 

We thus believe that we are at the beginning of a new era of great discoveries regarding the
solar interior. With the advent of powerful supercomputers we  start to non linearly
study the solar radiative interior and tachocline in full 3--D MHD simulations for the first time 
but lack the important and more detailed observational constraints that are required to 
progress in our understanding of the magnetohydrodynamics of the solar radiative interior.
A space project constraining the solar radiative dynamics is
clearly required to support and guide our 3-D simulations.  

We here briefly summarize some simulations of Brun \& Zahn (2005) and the dynamical properties that
we should look for in the next generation of solar seismic probes dedicated to the solar core.
The ASH code (Clune et al. 1999,
Brun et al. 2004) is used to model the solar radiative interior from $r=0.3$ to $r=0.71 R_\odot$ , 
assuming a stable stratification throughout the computational 
domain deduced from the 1--D solar model (Brun et al. 2002). 


Figure 3 displays the temporal evolution of the angular velocity $\Omega$ for one model of 
the solar radiative interior, with the poloidal field lines superimposed in black. 
The left panels represent both the initial solid body rotation state 
of the radiative zone and the purely poloidal field (here deeply confined) used to 
start the temporal evolution. We see on the middle panels that the
imposed shear starts to propagate in the radiative interior at rather high latitudes.
After about 4 Gyr, the angular velocity profile in the radiative interior is not more solid. 
Wherever the field lines cross the top boundary, the imposed shear has propagated all the way down 
to the bottom of the domain. 

It is interesting to note that helioseismic inversions of the rotation of the radiative interior 
infer an almost flat profile (solid body rotation down to 0.2 R$_\odot$) that
rotates to a value very close to the value of 441 nHz which balances torques at the interface of the convective zone (Gilman et al. (1989). 
This means that the exchange
through the interface between the convection and the radiation zone are nearly balanced, with
only a weak torque operating there. How the angular momentum lost through the solar wind and resulting in
the spin down of the Sun influences the redistribution of angular momentum in the radiative interior 
and in the tachocline, constitutes a major puzzle that needs to be answered.
The simulation also reveals the presence of a weak multi-cells meridional circulation per
hemisphere in the tachocline, directed poleward at the top of the domain. 

These simulations allow to follow  the temporal evolution of the kinetic and magnetic energies (decomposed
in axisymmetric and non axisymmetric components). We note a strong amplification of the mean 
toroidal magnetic field (TME) by the so-called $\omega$-effect. Even more interesting, 
the non-axisymmetric kinetic and magnetic energies (FKE \& FME) grow by many order of magnitudes.
This has to be expected, since following Tayler (1973), we know that a purely poloidal field is
unstable. In the simulation, the magnetic field has evolved from a purely dipolar
configuration to a mixed poloidal/toroidal configuration, those strengths are about the same,
thus confirming Tayler's theoretical work (see also Brainwaith \& Spruit 2004).
 
It is important  to probe the solar interior in order to put some constraints on these models
and try to estimate if such complex/mixed poloidal-toroidal 
magnetic field configurations are realized.

It appears that a fossil dipole field is not that efficient at limiting the 
inward thermal hyperdiffusion of a latitudinal shear, once its field lines have connected
with the region where the shear is imposed, unless we bury it deeply enough. 
By burying the initial seed field deep enough we succeed at limiting the spread of the 
tachocline at low latitude. Whether in the real Sun, the fossil field lines connect or 
not with the convection zone is another challenging question to be answered.

\section{The open questions for observers}
We now list
some fundamental questions which will greatly help the MHD simulations and constitute 
a challenge for observers. We will consider three parts: the knowledge of the Sun itself, 
 the Sun for stellar physics, the Sun for fundamental physics.

\vspace {3mm}
\noindent
{\it The Sun is our star}

\noindent
The future instruments must be prepared to answer to the following 
questions: 
\begin{itemize}
\item what is the rotation in the solar nuclear core, is there a latitudinal dependence ?
\item is there a relic of the formation of the solar system: is there a higher rotation profile in the core ?
\item can we see magnetic field effects coming from the core and could we follow potential
variabilities with time ?
\item could we quantify the real present energy balance including the redistribution along time between different sources 
of energy mentioned in Fig 4, a more serious transport of photons across the star and finally get the proper sound speed profile. 
\end{itemize}
It is only after having established such a complete representation of the Sun, that we will be able to predict 
the slow luminosity variations along human 
scales and answer to the questions connected to the observation of great minima, the Wolf, Sporer, Maunder minima:
\begin{itemize}
\item Can we may imagine several kinds of dynamos, even in the radiative zone, several cycles ?
\item What is the interconnection of the radiative magnetic field to the convective one ? 
\end{itemize}
3D MHD simulations have to be guided by more observations of the solar radiative zone to determine 
which magnetic field configurations act in the solar radiative interior.

\vspace {3mm}
\noindent
  {\it The Sun for Astrophysics}
  
  \noindent
The Sun is a unique star where million modes can be detected, it allows detailed comparisons with simulations:
\begin{itemize}
\item the low degree modes 
are the only modes we may detect for other stars, so any technical improvement 
will be  generalized to other stars,
\item the Sun has always stimulated stellar advances, it stays the best case to 
check the new theoretical assumptions, in parallel 
to the growing asteroseismogy which multiplies the physical sites (COROT launch 2006 and "Life and Stars" proposal 
for the present Cosmic Vision). The Sun is a key to build 
an unified view of the stars. Effectively, if we understand the internal magnetic
field of the Sun and stimulate complete MHD calculations, we will get
a complete renewal of stellar evolution.  The magnetic effects that we lack in present stellar evolution
appear much more important in the first and final stages, so  the same 
equations will describe very properly all the stars from their formation to their death. 
\end{itemize}
The complete MHD vision of stars is fundamental to understand the
magnetic relationship between stars and planets, the SNIA physics for 
cosmological perspectives and massive supernovae for enrichment of galaxies. 

\vspace {3mm}
\noindent
 {\it The Sun and fundamental physics}
 
\noindent
Determining the central part of the Sun by helioseismology has been determinant to estimate  
the neutrino fluxes emitted by the Sun 
independently of constant new advances in the ingredients of the solar models: we can quote today 
abundance redetermination (Asplund et al. 2002): carbon oxygen, neon may be ...,
provoked by a better description of the turbulent atmosphere. After thirty years of experimental progress, 
we have reached a level where a perfect agreement within the two probes (waves and neutrinos)
is  established (Turck-Chi\`eze 2004b,2005b). This example
illustrates
the importance of observational facts for improving our solar knowledge. 

The seismic probe
will continue to inform us on the more complex physics of the radiative zone 
and will put stronger limits 
on dark matter (Lopes, Bertone \& Silk 2002 and references therein).

In the present context it is also interesting to notice that
the successive solar gravitational moments that determine the solar moments of inertia, 
are still poorly known. However, these moments tell 
us how much the Sun's material contents deviate from a purely spherical distribution 
and how much the velocity rate differs from a uniform distribution. Thus, their precise 
determinations give indications on the solar internal structure.
The dynamic study of the gravitational moments until now is mainly based on solar observations 
(helioseismology, solar diameter) and solar models of rotation and density (Pijpers 1998, Roxburgh 2001). Various methods 
(stellar structure equations coupled to a model of differential rotation, theory of the Figure of the Sun, 
helioseismology) lead to different estimates of $J_n$, which, if they agree on the order of magnitude, 
diverge for their precise values (Rozelot et al. 2004, a, b, c).

\begin{figure*}[t!]
\centering
\includegraphics[width=.39\textwidth,height=0.26\textheight]{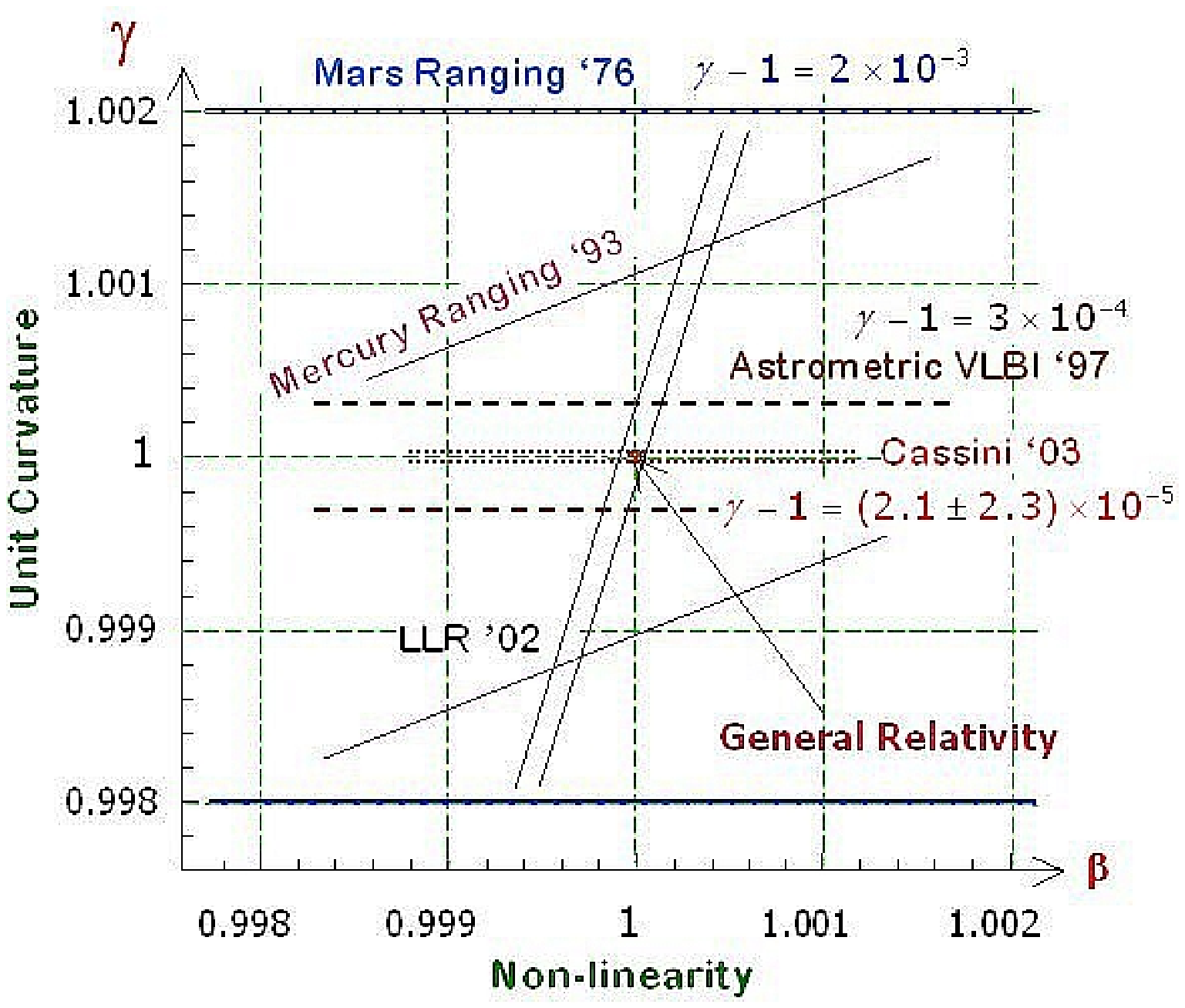}
\includegraphics[width=.39\textwidth,height=0.26\textheight]{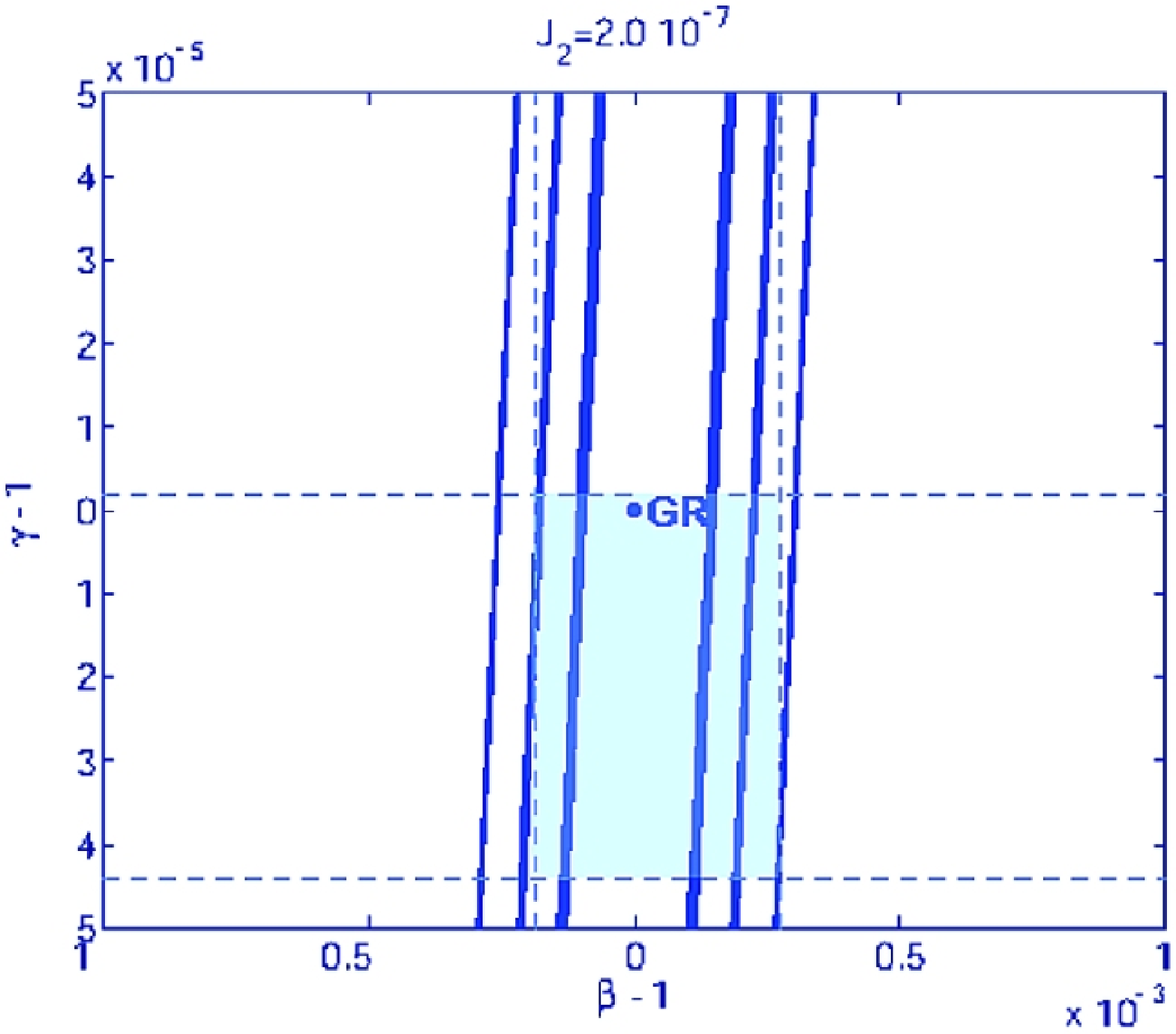}
         \caption{\small {Left (a): Thirty years of testing General Relativity from space. Right (b): A given value of $J_2$ constitutes a test of the PN parameters $\beta$ and $\gamma$. In the $\beta$ and $\gamma$ plane, the 1$\sigma$ (the smallest), 2$\sigma$ and 3$\sigma$ (the largest) confidence level ellipses are plotted. Those are based on the values for the observed perihelion advance of Mercury, $\Delta_{wobs}$, given in the literature and summarized in Pireaux and Rozelot (2003). General Relativity is still in the 3$\sigma$ contours for the allowed theoretical values of $J_2$ quoted on the upper part of the chart.}}
         \label{relativite}
\end{figure*}

If, from a physical point of view, these multipolar moments lead  to distortions of the solar surface, 
called asphericities, they also  have a dynamic role in the light deflection or in celestial mechanics. 
In the ephemerids computation, the determination of $J_2$ is strongly correlated with the determination of 
the Post-Newtonian Parameters (PPN)characterizing the relativistic theories of the gravitation. 
Lastly, the ignorance of $J_2$ is also a barrier to the determination of models of evolution of the solar system
on the long term.

The relativistic aspects are crucial in the dynamic approach of the solar parameters and open interesting 
pros- pects for the future. In this context, it is interesting to obtain a dynamic constraint of $J_2$, 
independent of the solar models of rotation and density, being used thereafter to force the solar models. 
Such a study is relevant in the scope of  space missions such as BeppiColombo (better determination of the PPN; 
possible measurement of the precession of the apside line of Mercury as a function of $J_2$), GAIA (better 
determination of the PPN, possible decorrelation PPN--$J_2$  thanks to the relativistic advance of the 
perihelion of planets and minor planets) and obviously the space version of GOLF-NG or DYNAMICS (precise determination of the 
rotation of the core where half of the mass is concentrated). Another key parameter 
of the solar models, which could also be constrained in a dynamic way is its spin, from the spin-orbit 
couplings which is introduced in celestial mechanics. From present Solar System experiments 
(Lunar Laser Ranging, Cassini Doppler experiments, etc, see Pireaux and Rozelot, 2003), it turns 
out that General Relativity is not excluded by those, as shown in the most up to date values in Fig. 
\ref{relativite}a. However, General Relativity would be incompatible with the Mercury perihelion advance 
test if $J_2$ = 0 was assumed. But with a non zero $J_2$, General Relativity agrees with this latter test, 
and there is still room for an alternative theory too (see Fig. \ref{relativite}b). The prepared next generation of instruments
 should 
provide the necessary $J_n$ measurements.

\section{Gravity modes and the core dynamics}

The solar core dynamics has a too small effect on acoustic modes to determine the rotation profile 
of the solar nuclear core without ambiguity. Effectively these acoustic modes are mainly 
sensitive to the differential rotation of the superficial layers (Garcia et al. 2001, Couvidat et al. 2003a). 
On the contrary, gravity modes are mainly sensitive to 
this central region and must reveal different aspects of the deep internal dynamics if they are detected.
They have been searched for thirty years.

But these modes produce very small intensity fluctuations at the surface or velocities of probably 
no more than fraction of mm/s (Kumar et al., 1996).
The first limit from ground observations was 7 cm/s (Delache \& Scherrer 1983, van de Raay 1988). 
Then the limit has decreased to 1 cm/s for single peaks 
(Appourchaux et al. 2000), a limit which has been 
reduced to 6 mm/s  for the GOLF instrument after two years of observation (Gabriel et al. 2002). 

In fact, the only hope to detect some gravity modes with the present observing instruments and to learn 
some information on the dynamics of the core is to look for patterns in the low frequency asymptotic region 
after a very long and 
continuous observation time (Garcia et al. in preparation) or to look for multiplets in the higher part of 
the gravity mode
spectrum, that means in the region of mixed modes. These modes have their amplitudes greater by 
one order of magnitude than the other 
gravity modes. 

\begin{figure*}[t!]
\centering
\includegraphics[width=.39\textwidth,height=0.26\textheight]{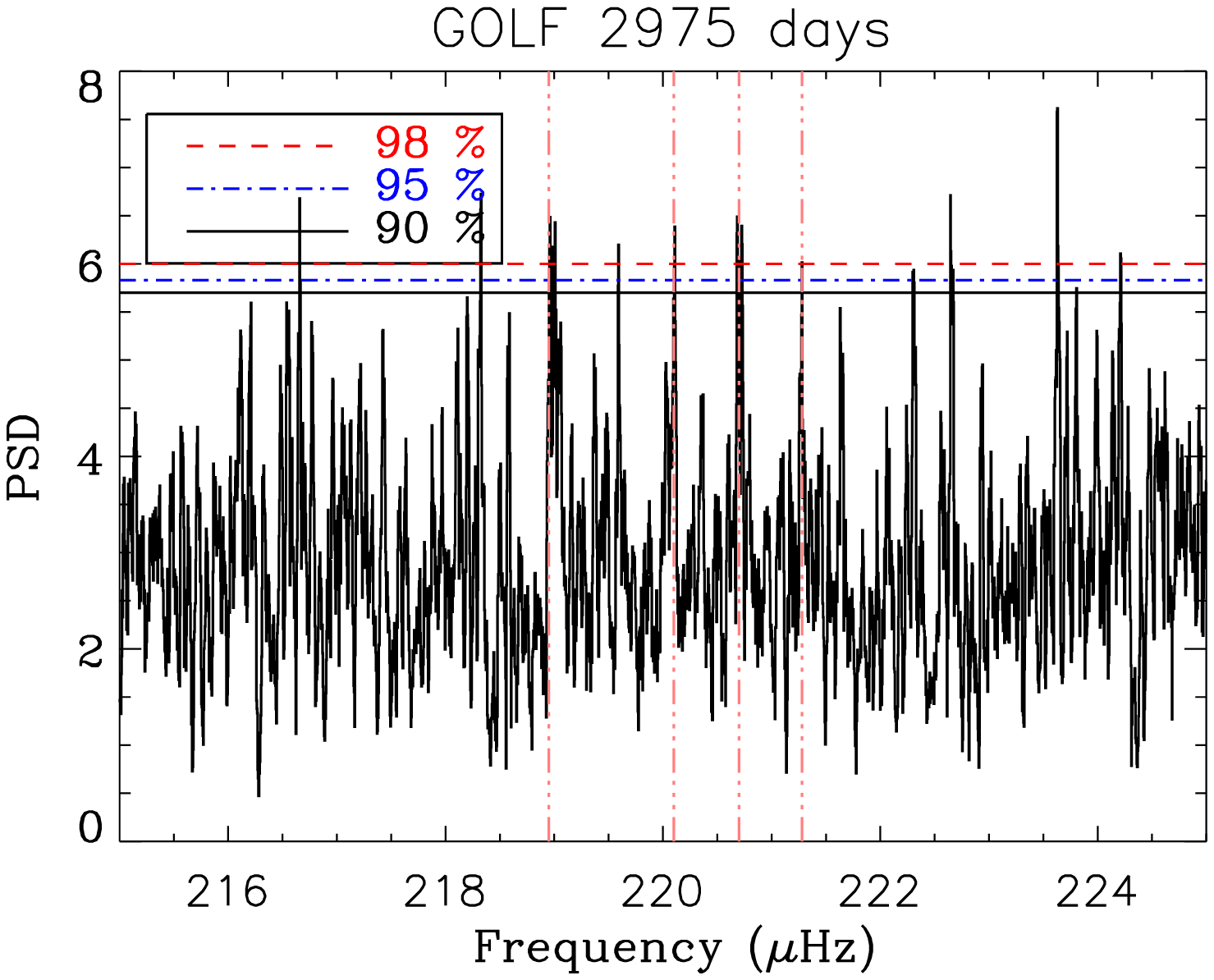}
\includegraphics[width=.39\textwidth,height=0.26\textheight]{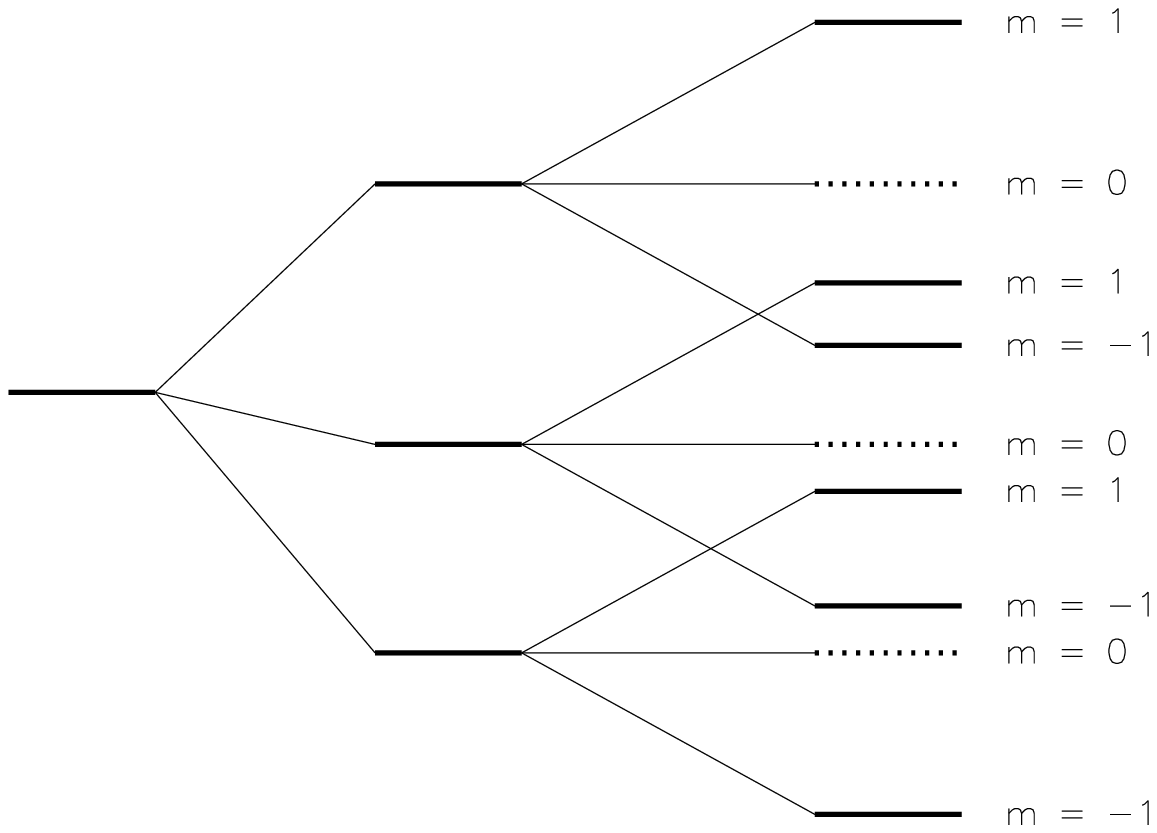}
         \caption{\small {Left (a): Interesting quintuplet detected in GOLF/SoHO with more than 98\%
confidence level after 2975 days of observation (Turck-Chi\`eze et al. 2004a, c); 
Right (b) Schematic of the origin of an hyperfine splitting for an $l=1$ mode. 
In the non-rotating, non-magnetic case there is a single frequency
(left). In a corotating frame, rotation and magnetic field split
the frequency, but each of the three normal modes is, in the
non-axisymmetric case, a mixture of different $m$ values (middle).
Transformed into the observer's frame, each $m$ value is shifted
and gives rise to hyperfine splitting (right). Dotted thick lines
denote those components that would not be visible 
from the ecliptic plane in whole-disk measurements.}}
\label{gravity}
\end{figure*}
Following this second idea, the analysis of the GOLF instrument, built with a very low intrinsic noise,  
has fixed the detection limit for the mixed modes to 1 mm/s (Turck-Chi\`eze et al. 2004a). At that limit,
interesting patterns have emerged as quintuplets or sextuplets and have been followed in time. 
It is interesting to note the presence of multiplets containing five peaks (instead 2 or 3), after 2975 days
 of observation, with more than 98\% confidence level not to be pure noise in 10 $\mu Hz$
(Fig 6a, Turck-Chi\`eze et al. 2004a, c). Of course these patterns
 may be noise, but they can also reveal some interesting physics as mixing between adjacent modes, or
an increased central rotation
(as shown in Fig 1b) and some signature of internal magnetic field (Fig 6b). 
The two persistent cases (fig 10 of Turck-Chi\`eze et al. 2004a) correspond to two regions where an $\ell = 2$ ($\ell = 1$) 
are seated on an $\ell = 5$ ($\ell = 4$). The presence of different modes at the same frequency may lead to an 
amplification of the signal (Cox \& Guzik 2004).
Moreover the lifetime of these modes is now predicted as considerably reduced and reexcitation is possible (Dintrans et al. 2005). 

In fact, the frequencies of the sun's acoustic (p) and gravity (g) modes 
are sensitive to the rotation and magnetism of the solar interior. 
Rotation, magnetism and other asphericities 
raise a degeneracy in the frequencies, leading to frequency 
splitting which has been
used by helioseismology to map the internal rotation of much of the 
solar interior. The greater sensitivity of g modes to the solar
core, compared with p-mode sensitivities, raises the possibility 
of detecting `unexpected' core properties, such as rotation on an
axis different from that of the solar envelope, or an oblique 
core magnetic field.

A global mode is characterized by its
radial order $n$, degree $l$ and azimuthal order $m$. If the rotation 
and magnetic field are axisymmetric $m$ is a well-defined quantum 
number and frequency splitting gives rise to up to $2l+1$ observable
frequencies for given $n$ and $l$ (i.e. $m=-l,\ldots,l$). The 
frequency splitting can then be decomposed into a part that is an odd 
function of $m$ and another that is an even function. The odd
component arises from rotation, and can be used to constrain that,
whilst the internal magnetic field contributes to the even 
component.  

If magnetic field breaks the axisymmetry, e.g. an inclined field, then each 
eigenfunction is a mixture of $m$ values. This leads to the so-called
hyperfine splitting: in principle up to $(2l+1)^2$ frequencies
can be apparent in the observer's frame (figure 6b). Such hyperfine splitting
may already have been detected for a g-mode multiplet (Turck-Chi\`eze
et al. 2004a, figure 6a). 

This idea has been applied for inclined radiative zone rotation
and for an inclined magnetic field of a relic field (Goode \& Thompson, 1992).
For spatially unresolved whole-disk measurements (such as GOLF), in the
axisymmetric case only $l+1$ components are visible (i.e. $l+m$ even). 
In the case of non-axisymmetric steady rotation, up to $(2l+1)(l+1)$
components are visible. How many of these are actually at visible
amplitude depends on the mix of $m$ values in each normal mode in the 
corotating frame. If unresolved, such hyperfine splitting may
appear as excess mode linewidths.
Based on published numerical calculations for p modes, one
can estimate that in spite of their larger Ledoux factors, the 
hyperfine splitting of g modes can be dominated by a $10^8$ gauss
magnetic field in the Sun's core, producing splittings of as much as 
5 $\mu$Hz. 

The idea that the solar core includes a potential relic of the planetary system formation is very exciting. SoHO observations will continue for 
two years but the quality of observation must be maintained.
Already very few identified g-modes will largely improve our knowledge of the solar core. 
In parallel we need
to improve the technique of detection which has been extended from ground networks.
The objectives are to measure more quickly (typically every year) all the
low-degree p modes
and the g-modes up to $\ell= 5$ along at least two cycles as it is already the case for the solar luminosity and 
will be the case
soon for the 
solar deformations.

\section{Defining the roadmap}
We propose now a roadmap for Europe, in parallel to the American effort, to converge altogether to build 
a complete MHD view of the Sun 
from to the core to the corona. Let first describe quickly the already planned projects. 

In addition to STEREO, launch 2006, which will enrich our solar 3D vision for the external phenomena, 
the American project HMI/SDO (launch in 2008) will follow MDI/SoHO with an increased resolution.
This instrument will largely improve (1) our understanding of the convective zone dynamics and the related solar dynamo,
(2) the origin and evolution of sunspots, active regions and other indicators of activity, (3) the sources and drivers 
of solar activity or disturbances. This instrument does not include any improvement for gravity mode detection, 
the knowledge of the nuclear core is considered in this project as a secondary objective.

PICARD, a CNES microsatellite (launch mid 2008), 
will search for an increased visibility of the modes due to an amplification at the solar limb. The payload consists of two radiometers, 
three 4-channel sunphotometers and an imaging telescope on board a microsatellite placed in a sun-synchronous orbit. 
These instruments are under the responsibility of IRMB (B), PMOD (CH), and SA (F).The PICARD mission is a solar-climate 
relationship investigation. For that, its scientific objectives are the understanding the solar 
activity origin by measuring simultaneously several key parameters for solar modelling:
(1)	precise images of the Sun at several wavelengths to measure the solar diameter, 
asphericity, and limb shape and differential rotation as a function of the heliographic latitude,
(2) frequencies of oscillations,
(3)	measurements of the total solar irradiance and its variation,
(4)	spectral irradiance measurements relevant for the variation of the ozone concentration 
in the terrestrial atmosphere, and their variations as a function of time.
In addition, ground based correlative measurements of the solar diameter will allow to investigate 
the role of the atmosphere by comparing with measurements obtained with the same instrument placed in orbit.
The above measurements will allow the development of a solar and a climate model in which UV solar variability 
will be taken into account.
This mission is named PICARD after the pioneering work of Jean Picard (1620-1682) who determined precisely 
the solar diameter during the Maunder minimum.  
Apart this project, the European mission Solar Orbiter (launch 2013)
will dedicated its efforts to the external part of the Sun. An other project american 
Solar Polar Imager will look to the poles but will be also mainly focus on magnetic effects not too deep.

In parallel to these already programmed missions, two instruments are being developed to prepare a better 
signal/noise ratio for the detection of very low signals. These instruments have been proposed in 1999 
to ESA in the framework of the F2 call for mission 
as part of the SOLARIS project. This complete observatory of the Sun has been 
judged too big for this class mission. For these two cases, prototypes have been built to demonstrate their 
feasibility.

GOLF-NG, an improved resonant spectrometer, is a french-spanish initiative. It is based on the 
measurement of the Doppler velocity 
through 15 points, every 10s on the sodium D1 line and a measurement of the continuum. 
A prototype is in construction to study the hard subsystems: variable magnet, thermal and mechanics of the cell, 
stability of the detector. It
will be installed for ground tests in Tenerife in 2006. 
In this instrument, the mode velocities will be isolated 
from the magnetic perturbation of the line.
This improvement will reduce the solar granulation noise which dominates the g-mode region, 
(Espagnet et al. 1995, Garcia et al. 2004). The signal/noise ratio will be increased
by a factor at least 10 in the g-mode region (Turck-Chi\`eze et al. 2001c, 2005a) and 
the photon noise statistics by the same amount. It will explore the gravity mode region and
measure the variability of all the acoustic modes. The space version will contain different masks for the identification of the modes and will detect modes up to $\ell = 5$. An extremely 
precise time evolution of the sodium line, compared to a theoretical estimate of the line (Eibe et al.
2001) will also give magnetic information 
on the low atmosphere just above the photosphere. 
This instrument will benefit from previous works dedicated to
 the knowledge of the solar granulation, supergranulation and excitation of the modes (Roudier et al, 2001). 

Another new Italian-American instrument is being built to measure the velocity, intensity and line-of-sight 
magnetic fields of the full solar disk, simultaneously at four heights in the solar atmosphere with 
a resolution of 4 arc-seconds and a cadence of 10 seconds. The heart of the instrument is the 
magneto-optical filter (Cacciani \& Fofi, 1978) that can be operated using vapor cells containing K, Na, Ca and He. 
The instrument is also designed to have a high-resolution imaging mode that will provide 1 arc-second resolution 
over a FOV of 450x450 square arc-seconds. This instrument is scheduled for deployment to South Pole 
during the Austral summer of 2005/2006. 

These two complementary
techniques will improve our ability to detect gravity modes in space. 
Nobody believes today that it is possible to detect gravity 
modes from ground because of their very small signals,
and the atmospheric perturbation and the low duty cycle on ground.  
So,
we suggest and consider important to launch a small European mission we call "DynAMICS ": Dynamics and Magnetism 
of the Inner Core of the Sun, a microsatellite or a minisatellite just after
the launch of PICARD in order to maximize the science return during the next solar cycle,
  on the solar core region and on the atmosphere. 
This mission could include GOLF-NG and MOF instruments (or at least one of them).
With the mission DynaMICS in parallel to the other missions SDO and PICARD, we secure rapid progress on 
the dynamical effects from the core to about 600 km above the photosphere. Doing so and for
 a low supplementary cost, 
Europe will play one of the leading role in that important domain.  
  
The next two cycles are determinant to put 
constraints on the topology 
of the deep internal magnetic field and its evolution with time and to determine the whole complex mechanisms which provoke magnetic changes at the surface 
(luminosity, eruptive activity and may be different kind of cycles...).
 We cannot any longer ignore the dynamics of 98\% of the mass of the Sun 
contained in the radiative zone to describe the complete activity of the Sun. 
 All the useful techniques must be qualified during the next solar cycle.
 With such intermediate step, we will be in a very good position to send 
a world-class mission, near the 2018 solar minimum, with already selected instruments 
to follow all the phenomena during another solar cycle. The final "world-class" mission will benefit from
 an orbit around the L1 lagrangian point or a mission around the solar poles (see Beleinos proposal). 
In parallel, supercomputers will be largely improved. So we will be able to leave our simple 1D representation 
to go towards a complete 3D evolution of the Sun during the next solar cycle
with the necessary constraints coming from the central region of the Sun. 
The final mission will allow well mastered predictions 
of the solar activity on long (greater than the known 22 years) human timescales 
and their consequences for a complete understanding 
of the Sun-Earth connection 
applicable also
to other planetary systems.

So, the roadmap is composed of 3 steps, it guarantees sucessful results, 
with step by step 
progress and a maximal scientific return:
\begin{itemize}
\item ground networks, SoHO and TRACE (1990-2007) ar the first step 
of this solar dynamical perspective; 
\item then SDO, PICARD and a small European project, DyNAMICS during the next solar cycle 2008-2017, 
to obtain a detailed analysis of the the convective zone, and the tachocline, to follow the luminosity and radius 
variabilities, to qualify the stability of the instruments for the gravity mode search and to
measure constraints on the core dynamics. This instrumental effort will guided the "3D MHD model"
of our star.
\item and finally a world-class mission launched in 2018 near the solar poles or around the L1 Lagrangian point
to obtain all the necessary constraints to build
 a "predictable magnetic model" of the Sun and its relationship with our planet which we will extend to other stars and planets.
\end{itemize}

\begin{acknowledgements}
We would like to thank all our colleagues, engineers and technicians 
who collaborate with us and build extremely sophisticated instruments
for a best knowledge of our star. We would like also to thank our institutions and space national 
compagnies who help us to realize our dream.
  
\end{acknowledgements}

\end{document}